\documentclass[10pt,letterpaper,twocolumn,revtex4-1]{article} 

\usepackage{ol2}
\usepackage[draft]{hyperref}
\usepackage{amsmath}
\usepackage{epsfig}
\usepackage{graphics}
\usepackage{amssymb}
\begin{document}

\twocolumn[ 

\title{Self-stabilized Quantum Optical Fredkin Gate}


\author{Jonathan Hu,$^{1}$ Yu-Ping Huang,$^2$ and Prem Kumar$^2$}

\address{
$^1$Baylor University, One Bear Place \#97356, Waco, TX 76798 \\
$^2$Center for Photonic Communication and Computing, EECS Department, Northwestern University, Evanston, IL 60208\\
Email: jonathan\_hu@baylor.edu and yphuangpx@gmail.com
}

\begin{abstract} Quantum optical Fredkin gate is an indispensable resource for networkable quantum applications. Its performance in practical implementations, however, is limited fundamentally by the inherent quantum fluctuations of the pump waves. We demonstrate a method to overcome this drawback by exploiting stimulated Raman scattering in fiber-based implementations. Using a Sagnac fiber-loop switch as a specific example,  we show that high switching contrast can be maintained even in the presence of significant pump fluctuations. This unique feature of self-stabilization, together with high-speed and low-loss performance of such devices, point to a viable technology for practical quantum communications.
\end{abstract}

\ocis{270.5565, 130.4815, 120.5790.}

] 

\noindent All-optical information processing architectures, particularly
those exploiting quantum features of single photons,
promise significantly higher speed, lower energy
cost, and much lager capacities than existing electronic-based
designs~\cite{Miller_NP_4_2010,Nielsen_QCQI_2000}. Crucial to implementing
such architectures is the capability to route photonic
signals with low loss, low noise, and without disturbing
their quantum states. There exist two
basic types of switching devices that can potentially fulfill these
requirements. The first is based on a micro-cavity
design, utilizing the optically-induced quantum-Zeno effect~\cite{Jacobs_PRA_063830_2009,Huang_OL_2376_2010,Huang_JSTQE_2011}. The second type is a traveling-wave
design that exploits cross-phase modulation (XPM)
in a Fredkin-gate setup, or its derivatives~\cite{Fredkin_IJTP_219_2982,Milburn_PRL_2124,Mortimer_JLT_1217_1988,Blow_OL_248_1990}. Comparatively,
the Fredkin-gate switches do not require high-Q
optical cavities and can thus be more convenient and robust
for practical use. Recently, a Sagnac fiber-loop
implementation of such switches has been developed,
demonstrating ultrafast redirection of quantum-entangled photonic signals with low loss
and without introducing
any measurable degradation in the entanglement fidelity~\cite{Hall_PRL_053901_2011}.

The performance of all-optical switches of the Fredkin-gate design,
however, is fundamentally restricted by the
fluctuations in the power of the pump waves that drive the XPM process,
resulting in reduced switching contrast~\cite{Milburn_PRL_2124,Liebman_PRA_4528_1993}. To address this issue,
use of sub-Poissonian pump pulses with suppressed power fluctuations has been
proposed~\cite{Sanders_JOSAB_915_1992}. However, such pump pulses are hard
to generate experimentally. In this Letter, we propose
a new avenue to overcoming this fundamental difficulty
by exploiting the inherent nonlinearity of the optical medium constituting the Fredkin-gate switch.
Using a Sagnac fiber-loop switch as a concrete example~\cite{Hall_PRL_053901_2011}, we show
how self-stabilized switching can be obtained
even when significant pump fluctuations are present. As
a result, no precise control over the pump power is
needed while still achieving high switching contrast. Our theory
agrees with experimental data without the need for any
fitting parameter.

\begin{figure}
\centering
\epsfig{figure=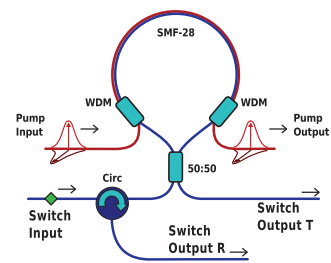,width=4.5cm}
      \caption{(Color online) A schematic of the fiber-loop switch based on the Kerr-nonlinear Sagnac effect. \label{Sagnac}}
\end{figure}

A schematic setup of the Kerr-nonlinear Sagnac fiber-loop switch is shown in Fig.~\ref{Sagnac}. Detailed description of such devices can be found in Refs.~13 and 14. Basically, it switches input signal waves between the transmission and reflection ports depending on the presence or absence of control pump waves.   In order to achieve polarization-insensitive switching, two orthogonally-polarized pumps of the same power are applied simultaneously~\cite{Hall_NJP_105004_2001}. The pumps are slightly detuned from each other within the correlation bandwidth of polarization-mode dispersion  in the fiber, so as to remain orthogonally polarized for the duration of XPM~\cite{Karlsson_OL_939_1999}. Under this condition, the signal experiences XPM effectively from a single polarized pump, and thus the entire system can be modeled with a pump wave that is co-polarized with the signal.

In our system, the pump pulse propagation can be described by the following generalized nonlinear Schr{\"{o}}dinger equation~(GNLS)~\cite{Agrawal_NFO_2006,Hu_OE_6722_2010},

   \vspace*{-0.1 in}
\begin{equation}
     \begin{array}{l}

     \displaystyle
\frac{\partial A(z,t)}{\partial z}+\frac{a}{2}A(z,t)-
\sum_{k\ge 2}\frac{i^{k+1}}{k!}\beta_k\frac{\partial^kA(z,t)}{\partial t^k}
=i \gamma \\
\displaystyle
	 	
		\times\left(1+\frac{i}{\omega_0}\frac{\partial}{\partial t}\right)\left[A(z,t)\int_{-\infty}^{t}R(t-t')|A(z,t')|^{2}dt'\right] ,
  \end{array}
  \label{GNLS}
\end{equation}

\noindent
where $A(z, t)$ is the electric-field envelope as a function of distance $z$ along the
fiber and retarded time $t$. The parameter $\omega_0$ is angular carrier frequency and $a$ describes the fiber loss, set as $a_Ae^{-a_B/\lambda} +
a_C/\lambda^4$ to include the infrared absorption and Rayleigh scattering, with $a_A = 5\times10^{11}$ dB/km, $a_B$ = 49 $\mu$m, and $a_C$ = 0.8
$\mu$m$^4\cdot$dB/km for SMF-28~\cite{Walker_JLT_1125_1986,Agrawal_NFO_2006}.
In Eq. (1), the parameter $\gamma=n_{2}\omega_0/cA_{\rm eff}(\omega)$ is the Kerr coefficient, where $n_2
= 2.2 \times 10^{12}$~m$^2$/W is the nonlinear refractive index~\cite{Agrawal_NFO_2006}, $c$ is the speed of light, and
$A_{\rm eff}(\omega)$ is the frequency-dependent effective area of the fiber, which is calculated for the linearly-polarized LP$_{01}$ mode of the SMF-28 with core diameter of 8.2~$\mu$m, cladding refractive index of 1.444, and core-cladding index difference of 0.36\%~\cite{Corning_SMF,Agrawal_NFO_2006}.
The nonlinear response function, $R(t)=(1-
f_{R})\delta(t)+f_{R}h_{R}(t)$, includes both the Kerr (instantaneous), $\delta(t)$,
contribution and the Raman (delayed), $h_R(t)$, contribution. We use the Raman response
function as $h_R(t) = \frac{\tau_1^2+\tau_2^2}{\tau_1\tau_2^2}
\exp(-t/\tau_2)\sin(t/\tau_1)$,
where $\tau_1 = 12.2$ fs and $\tau_2 = 32$ fs~\cite{Blow_JQE_2665_1989}. We set $f_R$ = 0.18 and 0 to simulate the cases with and without the
Raman effect, respectively. We use $D(\lambda)=S_0
(\lambda-\lambda_0^4/\lambda^3 )/4$ to approximate the dispersion in the SMF-28, where $S_0$ = 0.08 ps/(nm$^2$-km) and $\lambda_0 = 1313$ nm~\cite{Corning_SMF}.
From this dispersion formula, we obtain
the Taylor-series expansion coefficients at 1550 nm: $\beta_2 =19$ ps$^2$/km, $\beta_3$ = 0.11
ps$^3$/km, and $\beta_4=1.6\times 10^{4}$ ps$^4$/km. We found via simulation that the contribution of higher-order dispersion to the pump-pulse dynamics is negligible.


\begin{figure}
\centering
\epsfig{figure=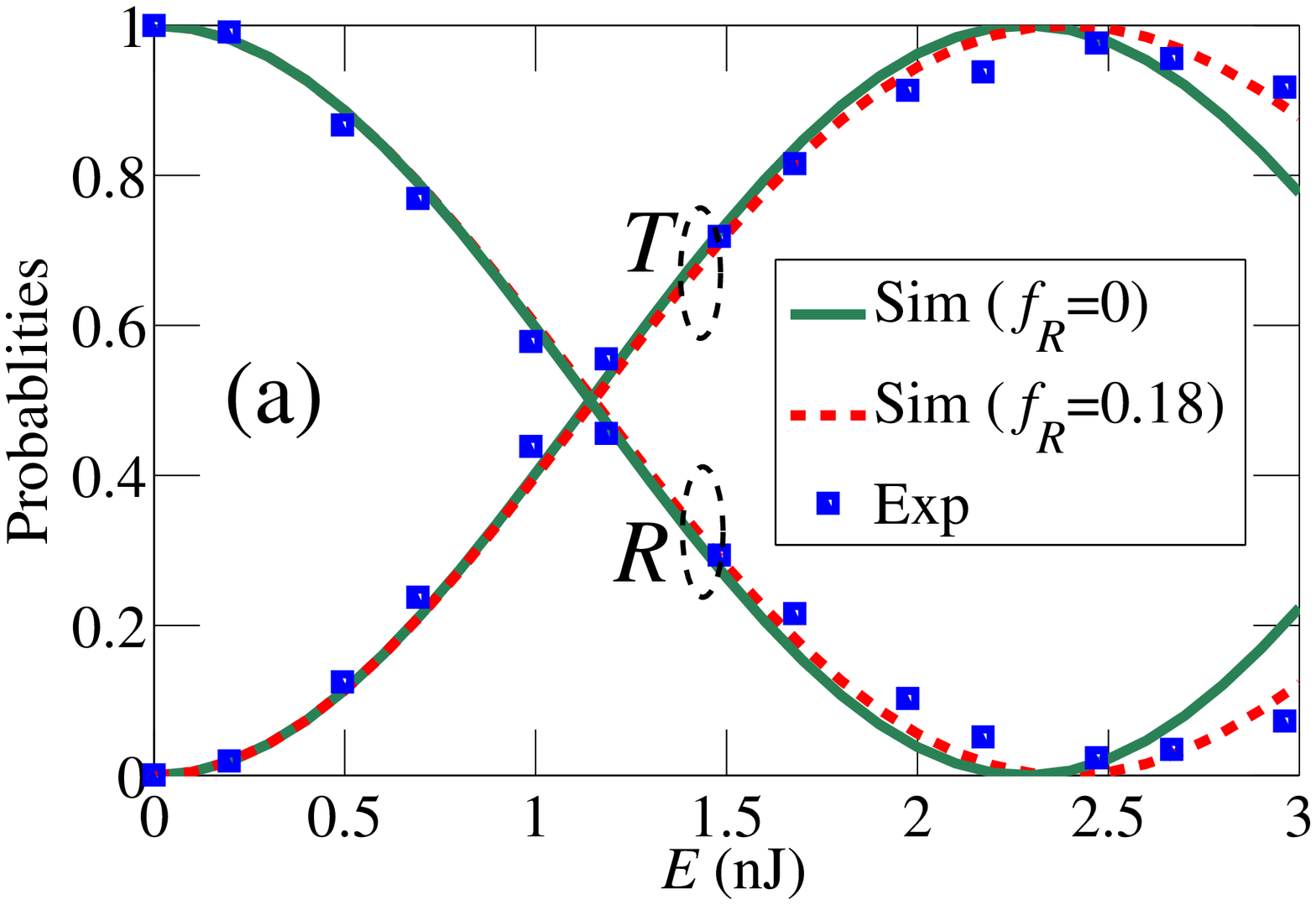,width=5.5cm}
  \epsfig{figure=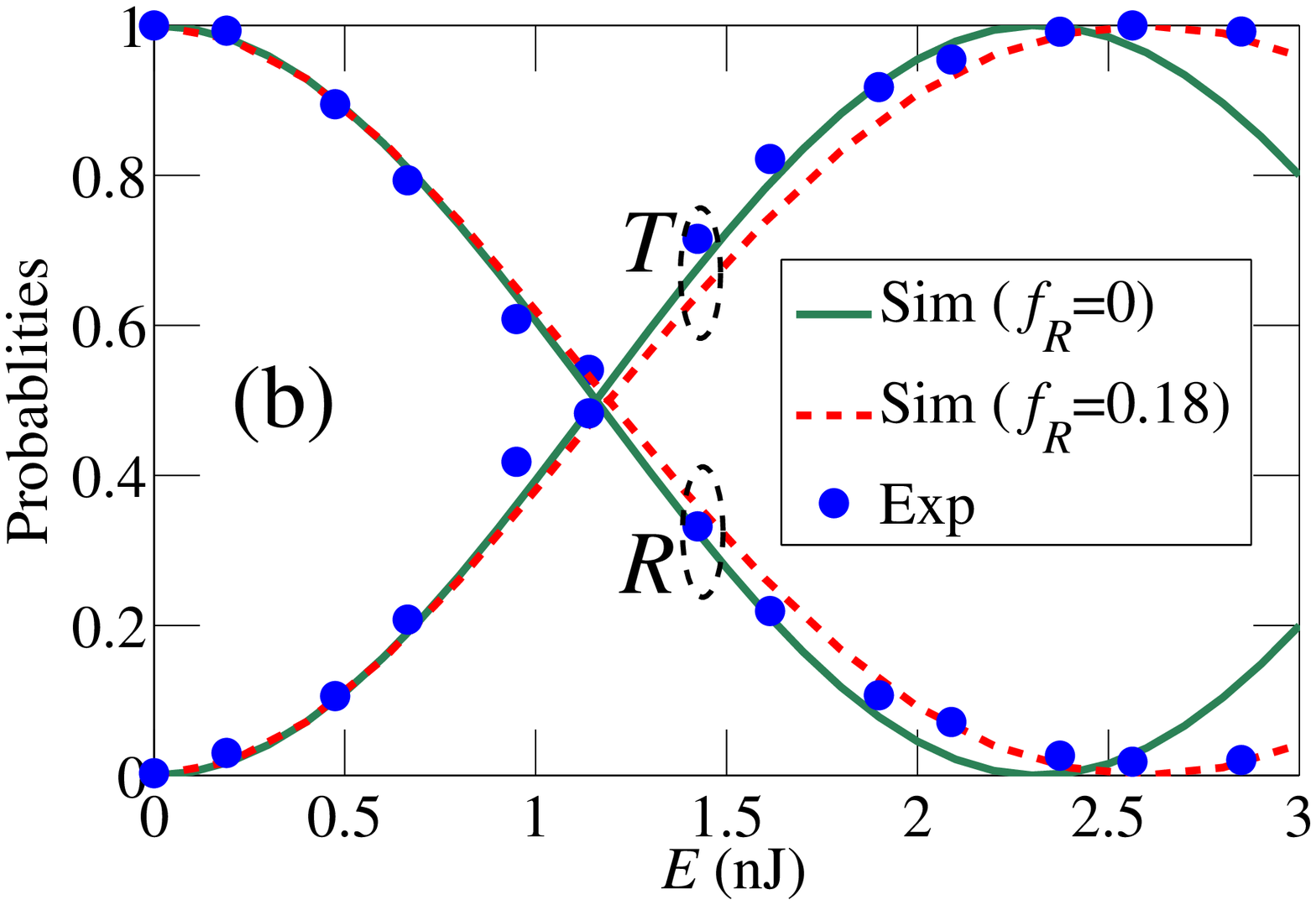,width=5.5cm}
      \caption{(Color online) Experimental and simulation results of peak switching probabilities vs. pump-pulse energy for fiber lengths of (a) 100 m and (b) 500 m. \label{TRE}}
\end{figure}

The phase shift induced by the pump is then calculated using the standard coupled equations~\cite{Blow_OL_248_1990}.
The normalized transmission ($T$) and reflection ($R$) are calculated using Eq. (8) of Ref. 14. Figures~\ref{TRE}(a) and \ref{TRE}(b) show the simulation
and experimental results with 5-ps full-width at half-maximum (FWHM) Gaussian pump pulses for fiber lengths of 100 m and 500 m, respectively. The red dashed
and green solid curves show the $T$ and $R$ with and without the Raman effect, respectively. As shown,
in both cases, the simulation results agree well with the experimental data at low pump-pulse energies. When the energy approaches 2.5 nJ, however, the
simulation results without the Raman effect diverge from the experimental data, predicting a sinusoidal behavior.
Taking the Raman effect into account, on the other hand, makes the simulation results to follow the experimental data throughout all the
pump-pulse energies we studied for both fiber lengths, showing that the switching probability flattens out as the pump-pulse
energy increases beyond the peak-switching point.

The different switching behaviors seen in Fig. 2 indicate clearly the important role of stimulated Raman scattering (SRS) in
our device. Indeed, in the region of high pump-pulse energies, SRS causes energy loss from the pump pulses by
shifting them to red-detuned wavelengths~\cite{Agrawal_NFO_2006}. Higher the pump-pulse energy, stronger is the SRS,
leading to correspondingly higher loss for the pump. Consequently, the XPM phase shift and the resulting switching contrast,
which are proportional to the pump-pulse energy, can remain almost constant because of the flattening out
behavior shown in Fig. 2. We note that such Raman effect cannot be interpreted as increased fiber loss at
longer wavelengths. In fact, our simulations show that even assuming wavelength-independent fiber loss, the same
saturating switching behavior persists.


\begin{figure}
\centering
\epsfig{figure=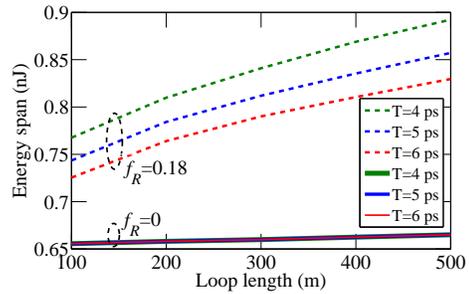,width=6.5cm}
      \caption{(Color online) Energy span for $T > 95$\% with (dashed) and without (solid) the Raman effect. The legend shows the different input pump-pulse widths. Note that the energy spans without the Raman effect are almost the same for all pulse widths. \label{SpanLength}}
\end{figure}

Because of the Raman effect, the switching contrast saturates at high pump energies, making the fiber-loop switch insensitive to
pump-energy fluctuations. Hence, there will be a wider range of the input pump-pulse energy that produces,
for example, $T > 95$\%.
For a fiber length of 100 m, the case shown in Fig. 2(a), the energy span is about 0.65 nJ, from 1.96 to 2.61 nJ,
without the Raman effect. With the Raman effect, in contrast, the energy span increases to 0.75 nJ,
from 2 to 2.75 nJ. Figure~\ref{SpanLength} shows the energy span for $T > 95$\% with and without
the Raman effect.
Green, blue, and red curves represent the energy span for input pump-pulse widths of 4 ps, 5 ps, and 6 ps, respectively.
Note that for a smaller pulse width, the peak pump power is larger for the same pulse energy, producing
a larger nonlinearity and SRS in pulse propagation. Hence the increase in the energy span due to the SRS
is more significant for the shorter pump pulses. According to our simulations, the energy span increases by
about 40\% for an input pump-pulse width of 4 ps and a loop length of 500 m, making the switching behavior more
insensitive to fluctuations in the pump-pulse energy. Even more significant improvements can be achieved for shorter pump pulses.

\begin{figure}
\centering
\epsfig{figure=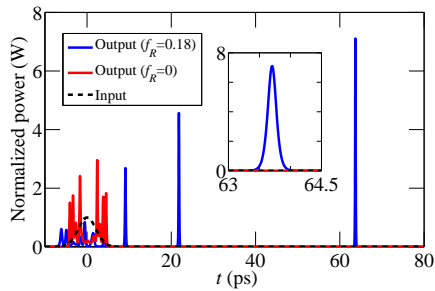,width=6.5cm}
      \caption{(Color online) Normalized power of the pump wave in time domain with a fiber length of 100 m. The inset shows the detailed pulse shape for the soliton in the time window between 63 and 64.5 ps. \label{PowerTime100}}
\end{figure}

\begin{figure}
\centering
\epsfig{figure=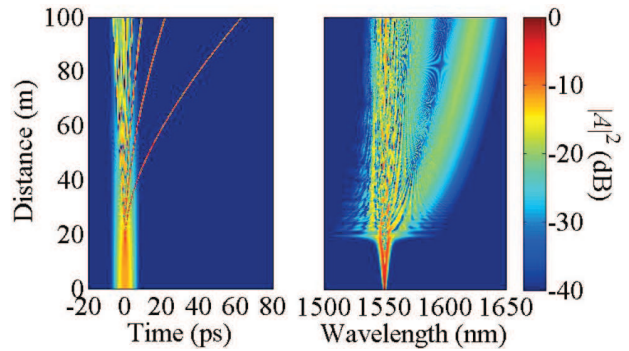,width=8.5cm}
      \caption{(Color online) Temporal and spectral
evolution during pump-pulse propagation, plotted on a log scale clipped at
$-40$ dB relative to the maximum. \label{ContTimeWave}}
\end{figure}

For a better understanding of the SRS effect, in Fig.~\ref{PowerTime100}, we show the normalized power of the pump wave in the time domain after propagating through a fiber length of 100 m. The red and blue curves represent the pump wave with and without the Raman effect, respectively. The black dashed curve shows the input pump pulse. We normalized the power in the plot such that the peak of the input pump pulse is 1~W. As shown, the pump pulse spreads out much wider with the Raman effect than without. This result is well explained by the temporal and spectral evolution of the pump pulses in the presence of SRS in propagation, as shown in Fig.~\ref{ContTimeWave}.
We note that within the first 10 m of fiber, the initial stage of propagation is
dominated by approximately symmetrical spectral
broadening due to self-phase modulation.
As a result, in the time domain, the pulse is compressed. Around the distance of 20 m, the spectral broadening becomes
strongly asymmetrical with the development of distinct
spectral peaks due to the soliton fission effect~\cite{Judge_OE_14960_2010}. The original pulse breaks into several
distinct individual solitons.
After the 20 m of propagation, the spectrum shows clear continuous redshift of the
long-wavelength components due to soliton self-frequency shift induced by the Raman effect~\cite{Gordon_OL_p662_1986}. In time, the pulses shifted to the longer wavelength also propagate slower than the pulses with shorter wavelength because of the anomalous group-velocity dispersion in the fiber. Hence, in Fig.~4, the simulation with the Raman effect shows pulses spreading into a wider time window.


In summary, we have demonstrated that by utilizing stimulated Raman scattering, a self-stabilizing quantum optical Fredkin gate can be realized in a fiber-loop setup. As a result, high switching contrast can be maintained without the need for precise control of the pump-pulse energy. Our theory agrees well with experimental data without the need for any fitting parameter. The theoretical and experimental results highlight the potential of a practical technology for implementing networkable quantum applications.


\vspace*{0.1 in}
J. Hu thanks C. R. Menyuk for useful discussions.

\vspace{-0.1 in}



\begin{thebibliography}{99}


\bibitem{Miller_NP_4_2010}
D. A. B. Miller, Nature Photonics {\bf 4}, 3 (2010).

\bibitem{Nielsen_QCQI_2000}
M. A. Nielsen and I. L. Chuang, Quantum Computation and Quantum Information (Cambridge University Press, 2000).

\bibitem{Jacobs_PRA_063830_2009}
B. C. Jacobs and J. D. Franson, Phys. Rev. A {\bf 79}, 063830
(2009).

\bibitem{Huang_OL_2376_2010}
Y. Huang and P. Kumar, Opt. Lett. {\bf 35}, 2376 (2010).

\bibitem{Huang_JSTQE_2011}
Y.-P. Huang and P. Kumar, IEEE J. Select. Topics Quantum Electron. {\bf 18}, 600 (2012).

\bibitem{Fredkin_IJTP_219_2982}
E. Fredkin and T. Toffoli, International Journal of Theoretical
Physics {\bf 21}, 219 (1982).

\bibitem{Milburn_PRL_2124}
G. J. Milburn, Phys. Rev. Lett. {\bf 62}, 2124 (1989).

\bibitem{Mortimer_JLT_1217_1988}
D. Mortimer, J. Lightwave Technol. {\bf 6}, 1217 (1989).

\bibitem{Blow_OL_248_1990}
K. J. Blow, N. J. Doran, B. K. Nayar, and B. P. Nelson, Opt. Lett. {\bf 15}, 248–-250 (1990).

\bibitem{Hall_PRL_053901_2011}
M. A. Hall, J. B. Altepeter, and P. Kumar, Phys. Rev.
Lett. {\bf 106}, 053901 (2011).

\bibitem{Liebman_PRA_4528_1993}
A. Liebman and G. J. Milburn, Phys. Rev. A {\bf 47}, 4528
(1993).

\bibitem{Sanders_JOSAB_915_1992}
B. C. Sanders and G. J. Milburn, J. Opt. Soc. Am. B {\bf 9},
915 (1992).

\bibitem{Hall_NJP_105004_2001}
M. A. Hall, J. B. Altepeter, and P. Kumar, New J. Phys. {\bf 13}, 105004 (2011).

\bibitem{Huang_NJP_053038_2012}
Y.-P. Huang and P. Kumar, New J. Phys. {\bf 14}, 053038 (2012).


\bibitem{Karlsson_OL_939_1999}
M. Karlsson and J. Brentel, Opt. Lett. {\bf 24}, 939 (1999).

\bibitem{Agrawal_NFO_2006}
G. P. Agrawal, Nonlinear Fiber Optics (Academic Press, New York, 2006).

\bibitem{Hu_OE_6722_2010}
J. Hu, C. R. Menyuk, L. B. Shaw, J. S. Sanghera, and I. D. Aggarwal, Opt. Express {\bf 18}, 6722 (2010).

\bibitem{Walker_JLT_1125_1986}
S. Walker, J. Lightwave Technol. {\bf 4} 1125 (1986).


\bibitem{Corning_SMF}
Corning SMF-28 Optical Fiber. Product Information Data sheet of Corning, Inc.


\bibitem{Blow_JQE_2665_1989}
K. J. Blow, D. Wood, IEEE J. Quantum Electron. {\bf 25}, 2665 (1989).



\bibitem{Judge_OE_14960_2010}
A. C. Judge, S. A. Dekker, R. Pant, C. M. de Sterke, and B. J. Eggleton, Opt. Express, {\bf 18}, 14960 (2010).

\bibitem{Gordon_OL_p662_1986}
J. P. Gordon, Opt. Lett. {\bf 11}, 662 (1986).



\end{thebibliography}
\end{document}